# Foundations for a Developmental State: A case for technical education

Professor Tshilidzi Marwala
Executive Dean: Faculty of Engineering and the Built Environment
University of Johannesburg
Auckland Park, Johannesburg, 2006
South Africa
Email: tmarwala@uj.ac.za

## Abstract

*This paper studies the viability of making a country a developmental state. In particular it studies the characteristics of a developmental state and how they are linked to technology. It then identifies technical education, as a vital force for the creation of a developmental state. In particular it identifies analytical, numeracy, computational and communication skills as vital forces for a society to create a developmental society.*

## Introduction
The goal of a democratic state is to establish a society in which the citizens are intellectually, socially, economically and politically empowered (Marwala, 2007b). In order to achieve this noble goal, certain conditions need to be in place in order to mobilize social, economic and political forces to capacitate the state to galvanize the productive forces that would ensure that these goals are achieved (Marwala, 2005b&c). One school of thought regarding the mechanism through which these productive forces can be galvanized, which is the subject of this paper, is to reorient the state as a developmental state (Chang, 1999). By so doing, sufficient productive forces will be unleashed to advance industrialization and this will principally require significant investment into technical education in primary, secondary and tertiary levels (Marwala, 2005a). This technical education should focus on creating a society that has sufficient analytical, numeracy, creative, computer and communication skills in society.

## Developmental State
A developmental state is a state where government is intimately involved in the macro- and micro-economic planning in order to grow the economy (Onis, 1991). It has generally been observed that successful developmental states are able to advance their economies much faster than regulatory states that use regulations to manage the economy. As an example, it took the USA approximately 50 years to double its economy while it took China, which is a developmental state, approximately 10 years to double its economy. Based on these findings, is it logical to infer that for a country to meet its social, economic and political obligations, it should become a developmental state? What are some of the characteristics that define a developmental state and is it possible to establish such characteristics in *South Africa*, which is case that would be considered in this paper?

## Characteristics of a Developmental State
In order to understand the concept of a developmental state, it is important to highlight some of the characteristics of a developmental state (Thompson, 1996; Woo-Cumings, 1999). Developmental states generally put strong emphasis on technical education and the development of numeracy and computer skills within the population. This technically oriented education is strategically used to capacitate government structures particularly the bureaucracy. What emerges out of this strategy is that the political and bureaucratic layers are populated by extremely educated people who have sufficient tools of analysis to be able to take leadership initiatives, based on sound scientific basis, at every level of decision making nodes within the government structure. Developmental states have been observed to be able to efficiently distribute and allocate resources and, therefore, invest optimally in critical areas that are the basis of industrialisation such as education. The complexity of the transformation agenda in South Africa makes the task of efficiently distributing and allocating resources difficult to achieve (Marwala, 2005c). The other characteristic that has been observed in successful developmental states is economic nationalism. This characteristic is also observed in developed states such as the USA during tough economic times. The characteristic of the national question in South Africa, which makes the notion of "South Africaness" a highly complex concept given the vast diversity of the South African population, makes economic nationalism not a viable option in South Africa. The other characteristic of a developmental state is its emphasis on market share over profit. The developed segment of the South African capitalist system is sophisticated and it has a huge component of short term investments also known as "hot money". This makes profit, particularly short term profit, a significant factor in the investment decision making process. Developmental states have been observed for their protection of their embryonic domestic industries and have also been observed to focus on aggressive acquisition of foreign technology. This they achieve by deploying their most talented students to overseas universities located in strategic and major centres of the innovation world and also by effectively utilizing their foreign missions (Marwala, 2005c; Marwala, 2006). Furthermore, they encourage and reward foreign companies that invest in building productive capacity such as manufacturing plants with the aim that the local industrial sector will in time be able to learn vital success factors from these companies. On constructing a harmonious social-industrial complex, developmental states strike a strategic alliance between the state, labour and industry in order to increase critical measures such as productivity, job security and industrial expansion. Even though developmental states do not create enemies unnecessarily and do not participate in the unnecessary criticism of countries with strategic technologies that they would like to acquire, they are, however, skeptical of copying foreign values without translating and infusing them with local characteristics. Developmental states generally believe that they will attain state legitimacy through delivery of services to citizens rather than through the ballot. In South Africa, state legitimacy is achieved through the ballot however the main shortcoming is that the society has not reached an equilibrium stage where the feedback mechanism between voting pattern and service delivering reinforce each other. Now that the characteristics of a developmental state have been highlighted it is important to briefly describe industrialisation because it is an important component of a developmental state.

**Industrialisation**
The vital driver for success in developmental states is industrialisation. The goal of industrialisation is to create a country that produces goods and services with high added values. For example, instead of exporting minerals unprocessed, people can be employed to beneficiate these minerals and manufacture goods such as watches and thus add economic value to the final products. The process by which countries add aggregate economic values to the products and services they offer is directly dependant on the level of industrialisation in the country's economy. The South African economy can be segmented into the so-called "two economies" where one part is highly industrialised and the other is underdeveloped. For South Africa to unify these "two economies" and tackle some of the serious problems it faces, it needs to build a developmental state whose foundations are outlined in the next section.

**Foundations for Building a Developmental State**
How does South Africa build a robust developmental state? What are the important characteristics of the industrial strategy that would get South Africa to advance at the fastest rate possible? What are the vital drivers in South Africa's social sphere that would accelerate development? On building a robust developmental state two aspects are vital and these are to vastly increase the level of educational attainment in the South African population and to increase the knowledge content in society particularly in the field of mathematics, science and computing. In particular, South Africa ought to vastly increase the numeracy and computer skills in the population and this can be achieved by introducing a robust early education strategy. This is because by the time the young learners go to school they have already acquired all the skills they require to develop numerical, computers and visualisation skills. It is vital that South Africa produces a cadre of highly educated people who are able to conduct advanced research and development to identify important areas of growth potential, plan the executions of the required solutions and monitor the implementation of the solutions proposed with a view of correcting the mistakes and reinforcing the successes. The vital characteristic of South Africa's industrial policy should be manufacturing but this should be synchronised with other key strategies such as rural development and agricultural policy. Since manufacturing is highly dependant on the productivity and the efficiency of the workers, it is vital that government, labour and industry reach a strategic pact that is focused on long term strategic goals rather than short term goals. As South Africa expands its agricultural output, particularly in rural areas, it ought to create local industrial centres where some of these agricultural products can be canned and preserved. South Africa ought to strengthen its co-operative strategies so that small businesses are able to efficiently integrate into the value chain between agricultural production and international markets. In summary, the foundation to building a developmental state is to develop: (1) an educated population with high levels of numeracy and computer skills; (2) a knowledgeable society with high levels of scientific literacy and appreciates the role of computers in building a knowledge economy; (3) a harmonious society with a strategic partnership amongst labour, government, industry and society; and (4) a society that efficiently allocates and distributes resources. In order to build this foundation it is important to pay a particular attention to engineering education which is the subject of the next section.

**Role of Engineering Education**
Engineering education is an enabling driver that is needed for South Africa to succeed as a developmental state. One of the main characteristics of successful developmental states is that they created an extensive bureaucratic layer consisting of mainly engineers who have high technical, computers and analytical skills (Marwala, 2007a). The need for planners who are highly numerate, computer-literate and posses highly developed visualisation skills, to be able to plan in large cycles that extend over long time periods, will require engineering schools to develop capacity to produce graduates with high competence in numeracy, visualization, computers and innovation. The need to produce technology workers that are able to adapt advanced manufacturing strategies to local settings requires broad engineering education that produces graduates that are highly competent in technical skills, analytical, computers as well as communication skills (Marwala, 2006). In order to achieve all these, South Africa ought to relook at the efficiency of the primary and secondary education so that the supply of students to engineering faculties is improved in both quality and quantity. Special attention ought to be paid to reconfiguring the social sphere so that the culture of appreciating the value of education is entrenched. In particular, technical education should be encouraged and the understanding that technical education is the driver of increasing developmental capacity. But more importantly South Africa needs to improve the subsidy funding formula to universities so that engineering education is efficiently funded to facilitate the attainment of the developmental expectations of the country. South Africa needs to internationalize the university system so that local students can be exposed to different and much more diverse ways of thinking and learning.

**Conclusion**
In conclusion the foundation for building a developmental state will be dependant on South Africa's ability to establish an educated population with high levels of numeracy as well as computational skills, creating a harmonious society with strategic partnerships amongst labour, government, industry and society as well as efficiently allocating and distributing resources.